\RequirePackage{snapshot} % required for bundeling as zip-file (see
\documentclass[10pt]{article}

\usepackage[disable]{todonotes}

\input{header}

\usepackage{booktabs} % For formal tables

\usepackage{microtype}
\usepackage{algorithmic}
\usepackage[ruled, vlined, linesnumbered]{algorithm2e}
\usepackage[utf8]{inputenc}
\usepackage{graphicx}
\usepackage{amsmath}
\usepackage{amsfonts}
\usepackage{amssymb}
\usepackage{balance}
\usepackage{array}
\usepackage{multirow}
\usepackage{tabularx}
\usepackage{subcaption}
\usepackage[justification=centering]{caption}
\usepackage{listings}

\usepackage{tikz}
\usetikzlibrary{calc}
\usetikzlibrary{intersections}
\usetikzlibrary{automata, arrows, positioning}

\tikzset{
  state/.style={
    rectangle,
    rounded corners,
    draw=black,
    minimum height=2em,
    inner sep=2pt,
    text centered,
    font=\normalfont,
  },
}
\tikzset{edge/.style={left, font={\sffamily\small}}}
\tikzstyle{line} = [draw, -latex']

\title{Functional Requirements-Based Automated Testing for Avionics}
\author{
  {Youcheng Sun\hspace{3mm} Martin Brain\hspace{3mm} Daniel Kroening}\\\bigskip
  {University of Oxford, UK}\\
  {Andrew Hawthorn\hspace{3mm} Thomas Wilson\hspace{3mm} Florian Schanda}\\\bigskip
  {Altran, UK}\\
  {Francisco Javier Guzmán Jiménez\hspace{3mm} Simon Daniel}\\\bigskip
  {Rolls-Royce, UK}\\
  {Chris Bryan\hspace{3mm} Ian Broster}\\
  {Rapita Systems, UK}
}

\date{}
\begin{document}

\maketitle

\begin{abstract}
   We propose and demonstrate a method for the reduction of testing effort
   in safety-critical software development using DO-178 guidance. 
   We~achieve this through the application of Bounded Model Checking (BMC)
   to formal low-level requirements, in order to generate tests
   automatically that are good enough to replace existing labor-intensive
   test writing procedures while maintaining independence from
   implementation artefacts.  Given that existing manual processes are often
   empirical and subjective, we begin by formally defining a metric, which
   extends recognized best practice from code coverage analysis strategies
   to generate tests that adequately cover the requirements.  We~then
   formulate the automated test generation procedure and apply its prototype
   in case studies with industrial partners.  In review, the method
   developed here is demonstrated to significantly reduce the human effort
   for the qualification of software products under DO-178 guidance.
\end{abstract}

%% The default list of authors is too long for headers}

%
%\keywords{...}

\section{Introduction}
\label{sec:introduction}

DO-178C/ED-12C~\cite{do178c} provides guidance for the production of
software for airbone systems.  For each process of the life-cycle it lists
the objectives for the life-cycle, the activities required to meet those
objectives and explains what evidence is required to demonstrate that the
objectives have been fulfilled.  DO-178C is the most recent version of the
DO-178 family.  DO-178 is thorough, and when the highest level is chosen,
aids the production of highly reliable software.

Developing to this guidance can be hard and time consuming.  We will use one
small aspect of this process as exemplar: the generation and verification of
low-level tests against low-level requirements (LLRs) as described in the
Software Reviews and Analyses part (Section 6 of the DO-178C).  The
objectives related to this particular area are: compliance with the low-level
requirements (6.3.4.a), compliance with the software architecture (6.3.4.b),
verifiability (6.3.4.c), conformance to standards (6.3.4.d), traceability
(6.3.4.e), accuracy and consistency (6.3.4.f), the executable object code
complies with the low-level requirements (6.4.c), the executable object code
is robust with the low-level requirements (6.4.d), the executable object
code is compatible with the target computer (6.4.e), normal range test cases
(4 additional objectives, 6.4.2.1), robustness test cases (6~additional
objectives, 6.4.2.2) and test coverage analysis (4~additional objectives,
6.4.4). A typical DO-178C test coverage analysis has two steps:
\begin{enumerate}

\item The first step is analysis of the coverage of requirements.

\item The second step is structural coverage analysis on the implementation.

\end{enumerate}
The requirements-based coverage analysis establishes whether the software
requirements are covered adequately, which may prompt a revision of the
requirements-based tests.  Subsequently, structural coverage analysis is
applied to determine which part of the code is exercised by the
requirements-based tests. Inadequate structural coverage in the second step
indicates that requirements are missing or that implementation behaviour is
unspecified.  To this end, requirements-based testing derives a suite of
tests from software requirements only, and must not use internal structure
of the implementation.

\paragraph{Structural Coverage Analysis}

The most common form of structural coverage analysis is \textbf{code
coverage analysis}, which measures the degree to which the source code of a
program has been covered during execution.  Criteria for code coverage
include statement coverage (checking whether each statement in the program
has been executed), branch coverage (checking whether each branch of
conditional structures in the code has been taken), and Modified
Condition/Decision coverage (MC/DC).  In MC/DC analysis, a boolean decision
consists of multiple boolean conditions such that every condition shall be
evaluated to true and false and it is required that this switch changes the
outcome of the final decision.  DO-178C guidance requires MC/DC coverage
of function bodies.

\paragraph{Automating functional requirements-based testing}

Common re\-quire\-ments-based testing techniques~\cite{bs7925} include
equivalence partitioning, boundary value analysis, decision tables, and state
transition testing.  Equivalence partitioning and boundary value analysis
are most relevant to the coverage criterion we propose.
Equivalence class partitioning is a software testing technique that partitions
each of the inputs (or outputs) to the unit under test into a finite number of
disjoint equivalence classes. It can also be applied to the inputs to a conditional
statement. It is usually used in conjunction with boundary value analysis. Boundary
value analysis is a technique that generates test cases that exercise an input or
predicate (conditional) variable just below, on, and just above the limits of valid
ranges (equivalence partitions), the rationale being that errors tend to occur near,
or on the partition boundaries.

The key goal of this paper is the automated creation of a functional test
suite from formal, low-level requirements.  
Software requirements are usually written in natural language and for our approach
these must then be translated into a formal specification in the form of pre- and 
post-conditions. 
The pre-conditions capture the calling context required before the function
call, and the post-conditions capture the state of the system required after
the function call.

There is a variety of approaches to automating specification-based
testing. In particular, \cite{gargantini1999using} is among first that
applied Model Checking to generate tests from requirements specifications. 
For each condition tested, a \emph{trap property} that violates this
condition is generated and instrumented into the code: if a trap property is
satisfied by the model checking, it means that the corresponding condition
will be never met by the program; otherwise, a counterexample will be
returned by the Model Checker that leads the program to the condition tested
and a test vector can be thus derived.  However, the method
in~\cite{gargantini1999using} needs to call the function body in a black box
fashion.  
In~\cite{chang1999structural}, structural
coverage is applied to function specifications to generate testing
conditions, which are then automatically matched with tests that already
exist.
\cite{hierons2009using} summarizes test selection strategies when using
formal specifications to support testing. In such cases,  
post-conditions are only used to determine the outcome of the
tests that are generated. Moreover, work like~\cite{tan2004specification}
and~\cite{whalen2006coverage} test specifications that include temporal
properties; though beyond the scope of the work in this paper, they
establish interesting future directions.

\paragraph{Main contributions.}
\begin{itemize}

\item We formally define a coverage criterion for generating testing
    conditions from software requirements.  It is based on ``MC/DC plus
    boundary analysis'', extended with control flow coverage.

\item We design an automated procedure for the test generation from pre- and
    post-conditions using a Model Checker.

\item We implemented the procedure in
  Model Checking tool CBMC~\cite{cbmc} and integrated it with
  RapiTestFramework (published by Rapita Systems). The resultant toolchain
  specializes in automatic testing under DO-178C and was applied in real-world
  case studies using source code for an avionics system provided by Rolls-Royce.
  We report here outcome from the case study work and experience learned from this project.

\end{itemize}

\paragraph{Structure of the paper}
In Section \ref{sec:problem-formulation}, we introduce the application
context for our automatic test case generation approach.
Section \ref{sec:requirements} gives the formal definition for
functional requirements, in form of pre- and post-conditions. In
Section \ref{sec:criterion}, a coverage criterion is formulated for
coverage analysis of requirements. Testing conditions
generated according to the proposed criterion, which are then utilized
by the automated testing framework developed in Section \ref{sec:auto-gen}
to generated test cases. The method introduced so far has been implemented
and integrated in a toolchain that is examined by two industrial case studies,
and this is reported in Section \ref{sec:case-study}. Finally,
we conclude the paper and discuss future exploitations in Section \ref{sec:conclusions}.

\section{The Problem Formulation}
\label{sec:problem-formulation}

\begin{figure}[!htb]
  \center
  \begin{tikzpicture}[scale=1.6]
  \node[draw=none, shape=rectangle, minimum width=1.6cm,minimum height=0.5cm] at (0, 5)   (hlr) {HLRs};
  \node[draw=none, shape=rectangle, minimum width=1.6cm,minimum height=0.5cm] at (0, 4)   (llr) {LLRs};
  \node[draw=none, shape=rectangle, minimum width=1.6cm,minimum height=0.5cm] at (-2, 3)   (testing) {Testing};
  \node[draw=none, shape=rectangle, minimum width=1.6cm,minimum height=0.5cm] at (2, 3)   (impl) {Impl};
  \node[draw=none, shape=rectangle, minimum width=1.6cm,minimum height=0.5cm] at (0, 2)   (val) {Validation};
  \node[draw=none, shape=rectangle, minimum width=1.6cm,minimum height=0.5cm] at (0, 1)   (evi) {Evidence};
  \node[draw=none, shape=rectangle] at (1.65, 2)   (text) {\small\textit{automation}};
  
  \node[draw, dashed, shape=rectangle, minimum width=3cm,minimum height=1.25cm] at (0, 2)   (box) {};

  \draw[->, thick] (hlr) -- (llr) node[midway,right] {\small \textit{human}};
  \draw[->, thick] (llr) -- (testing)  node[midway,left,yshift=0.2cm] {\color{red}\small\textit{human$\Rightarrow$automation}};
  \draw[->, thick] (llr) -- (impl)  node[midway,right,xshift=0cm, yshift=0.2cm] {\small\textit{human/automation}};
  \draw[->, thick] (testing) -- (val);
  \draw[->, thick] (impl) -- (val);
  \draw[->, thick] (val) -- (evi);

\end{tikzpicture}
\caption{The targeted software development scheme}
\label{fig:scenario}
\end{figure}
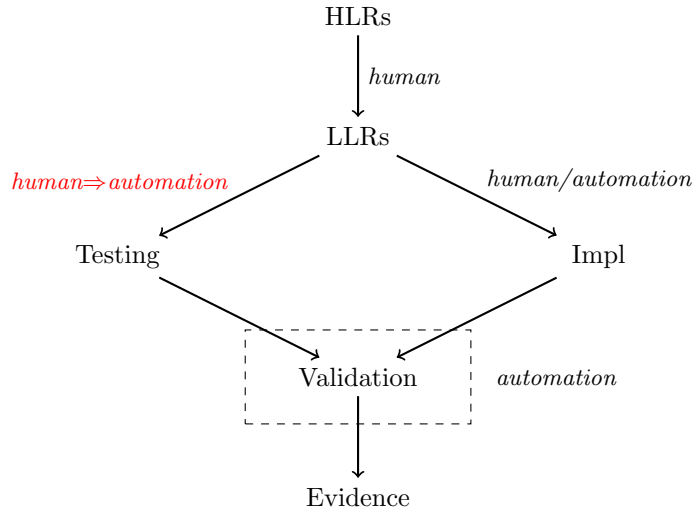

The work in this paper targets the scenario depicted in Figure
\ref{fig:scenario}, with our focus on using automated low-level requirements
(LLRs) testing to minimize human effort.
Within such a scheme, after high-level requirements (HLRs) are (mostly) manually
interpreted as low-level requirements (LLRs), the function implementation (Impl)
and the requirements testing are two independent procedures.
The function implementation can be either manually written by software programmers
or automatically generated by source code generation tools as in many model-based
development platforms. In the end, test cases generated from LLRs must be validated
against the function implementation (e.g., by code coverage level measurement),
which are supplied as evidence for fulfilling DO-178C guidance.

Figure \ref{fig:scenario} depicts a suitable scheme for developing
avionics software under DO-178C. The work in this paper specializes in the LLRs testing, which
is nowadays performed manually by experienced engineers. We aim to automating this 
procedure, that is, automatically creating test cases for LLRs.

Given the authors' experience in the field,
the cost of developing tests against low-level requirements (as required by
the objectives in DO-178C) equates to, on average, two days per
procedure/function.  This takes account of initial test development, test
debugging and review.  For a project with 250,000 lines of code and
assuming an average size of 25 lines per procedure/function, this equates to
50 person-years of effort!  Given this level of effort and expense, coupled
with the generally held view that low-level testing is of relatively low
value, there is a strong commercial incentive to consider ways of automating
this aspect of development.  The DO-333 Formal Methods
Supplement~\cite{do333} to RTCA DO-178C provides an approach to automating
low-level verification through the use of proof, but few businesses are
considering this option because of the perceived cost of using formal
methods. This is because verification results from a formal methods tool
(e.g., a Model Checker) can be used under DO-178C and DO-333
guidance, only if that tool has been certified or the verification process has
been instrumented to generate the evidence. Either way is challenging.

The research described in this paper has used formal methods in another way:
instead of mathematically proving the software, we use proof to generate
tests that cover low-level verification objectives (listed earlier)
through execution.  We expect (at least) the following advantages of this
approach.

\begin{itemize}
  \item \textit{Easy adoption} The automated tests represent a like-to-like
  replacement for existing (manual) test-writing procedures.  Generated tests
  can be reviewed in the same way as those written by a skilled engineer
  without a background in formal methods.
  \item \textit{Consistency} The usage of qualified tools that make use of
  formal methods improves consistency and completeness of testing by removing
  human variability.
  \item \textit{Improved development lifecycle} An automated process
  encourages developers to specify requirements at the beginning of the
  development lifecycle and to adopt test-driven practices.
  \item \textit{Reduced cost} As previously stated, a large percentage of the cost of safety-critical development is in low-level testing. Significant cost savings will be achieved by reducing or eliminating manual testing effort.
  \item \textit{Painless maintenance} If the code under test is changed, a revised set of tests can be generated without additional effort.
  \item \textit{Extensibility} When required, manual tests can be added alongside to complement generated tests in an identical format.
\end{itemize}

\section{The Requirement Specification}
\label{sec:requirements}

The software requirements must be formally specified in order to perform the
automatic test generation.  Works on specification languages
include~\cite{delahaye2013common} for C, \cite{barnett2004spec} for the
.NET, and SPARK Ada~\cite{spark}. Hierons et al.~\cite{hierons2009using} provide a
survey on formal specification techniques for testing.

Instead of digging into the details of some specific programming language
syntax, also for both the convenience and generality, we assume that every
requirement is specified as a Boolean expression \texttt{bool\_expr},
defined following the mini syntax in Listing \eqref{list:requirement}.

\lstset{language=C,
basicstyle=\ttfamily,
frame=single,
keywordstyle=\color{blue}\ttfamily,
stringstyle=\color{red}\ttfamily,
commentstyle=\color{green}\ttfamily,
morecomment=[l][\color{magenta}]{\#}
}
\begin{lstlisting}[mathescape=true, 
caption={The mini syntax of a requirement}, 
captionpos=b,
label={list:requirement}, 
%numbers=left,
commentstyle=\color{red}\ttfamily]
bool_expr $\stackrel{\text{def}}{=}$ true | false | bool_var | $\neg$bool_expr
            | bool_expr $\wedge$ bool_expr
            | bool_expr $\vee$ bool_expr
            | nbe $\{ <, \leq, =, \geq, >, \neq\}$ nbe
            | ITE(bool_expr, bool_expr, bool_expr)

non_bool_expr $\stackrel{\text{def}}{=}$ int_expr | float_expr
                | string_expr | char_expr | enum_expr
                | ITE(bool_expr, nbe, nbe)
\end{lstlisting}

%\lstset{language=C,
%basicstyle=\ttfamily, %\scriptsize,
%frame=single,
%keywordstyle=\color{blue}\ttfamily,
%stringstyle=\color{red}\ttfamily,
%commentstyle=\color{green}\ttfamily,
%morecomment=[l][\color{magenta}]{\#}
%}
%\begin{lstlisting}[mathescape=true, 
%float=*,
%caption={The mini syntax of a requirement}, 
%captionpos=b,
%label={list:requirement}, 
%%numbers=left,
%commentstyle=\color{red}\ttfamily]
%bool_expr $\stackrel{\text{def}}{=}$ true | false | bool_var | $\neg$bool_expr
%            | bool_expr $\wedge$ bool_expr
%            | bool_expr $\vee$ bool_expr
%            | non_bool_expr$\{ <, \leq, =, \geq, >, \neq\}$non_bool_expr
%            | ITE(bool_expr, bool_expr, bool_expr)
%
%non_bool_expr $\stackrel{\text{def}}{=}$ int_expr | float_expr
%                | string_expr | char_expr | enum_expr
%                | ITE(bool_expr, non_bool_expr, non_bool_expr)
%\end{lstlisting}

Most parts of the syntax are straightforward.  The
\texttt{ITE} is an if-then-else style operator like the ternary
operator ($?:$) in C language, and $nbe$ is short for $non\_bool\_expr$.
This mini syntax does not aim to be
comprehensive.  For example, function/procedure definitions are not specified here. 
However, an argument to the function call in the form of the Boolean
expression is treated as a \texttt{bool\_expr} and a function return is
regarded as a variable/expression of the corresponding type.
%
%In addition, when a function is called inside a requirement, boolean expressions within it are also covered.

In addition, a \emph{Boolean predicate} is defined as a Boolean expression
of the form $nbe \sim nbe$ with
$\sim\in\{<, \leq,\allowbreak =,\allowbreak \geq, >, \neq\}$.  A Boolean predicate
$nbe \sim nbe$ is also called
\emph{an ordered predicate} if $\sim\in\{<, =, >\}$.  A~\emph{compound
Boolean expression} is with multiple conjunctions or disjunctions of its
Boolean sub-expressions.

%\begin{equation}
%  \begin{aligned}
%    bool\mbox{-}expr&::=true|false| \neg bool\mbox{-}expr|bool\mbox{-}expr\wedge bool\mbox{-}expr| bool\mbox{-}expr\vee bool\mbox{-}expr\\
%                    &\mbox{   } | non\mbox{-}bool\mbox{-}expr \{ <, \leq, =, \geq, >, \neq\} non\mbox{-}bool\mbox{-}expr \\
%                    &\mbox{   } | ITE( bool\mbox{-}expr,  bool\mbox{-}expr,  bool\mbox{-}expr) \\
%                    &\mbox{   } | bool\mbox{-}return\mbox{-}function\mbox{-}call \\
%                    %&\mbox{   } | forall\mbox{-}expr | exists\mbox{-}expr \\
%    non\mbox{-}bool\mbox{-}expr&::=int\mbox{-}type|float\mbox{-}type|string\mbox{-}type|char\mbox{-}type |enum\mbox{-}type\\
%                               &\mbox{   } | ITE( bool\mbox{-}expr,  non\mbox{-}bool\mbox{-}expr,  non\mbox{-}bool\mbox{-}expr)\\
%                               &\mbox{   } | non\mbox{-}bool\mbox{-}return\mbox{-}function\mbox{-}call \\
%  \end{aligned}
%  \label{eq:spec}
%\end{equation}

\section{The Coverage Criterion}
\label{sec:criterion}

We assume that the requirements are given as \texttt{boo\_expr} as specified
in Listing \ref{list:requirement}.  We now develop a coverage criterion in
order to automatically apply a formalized procedure to each expression which
generates a set of testing conditions that meaningfully interpret this
requirement.  The coverage criterion in our work extends MC/DC with boundary
value analysis (for compound expression analysis) to take control flow
structures in the requirement specification into account.

\subsection{The compound expression}

Our treatment of compound expressions is similar
to~\cite{chilenski2001investigation}, which uses MC/DC together with
boundary value analysis.  However, we provide a formal definition of the
rules for generating the testing conditions.

In code coverage analysis, the MC/DC criterion is often regarded as a good
practice to generate testing conditions to examine a compound expression. 
According to MC/DC, the Boolean expression being tested is called the
\emph{decision}, which is going to be evaluated to true or false, and a
\emph{condition} of the decision is a Boolean predicate that cannot be
broken down into simpler Boolean expressions.  The MC/DC criterion requires
that each condition needs to be evaluated one time to true and one time to
false, and this flip of condition value affects the decision's output.  For
example, let us consider a requirement specified as in following compound
expression.
\begin{equation}
  \label{eq:mcdc-example}
   M\geq 0\wedge N\leq1000
\end{equation}
Applying the MC/DC rules, we obtain the testing conditions
$\Phi=\{\allowbreak M\geq 0\wedge N\leq1000,\allowbreak
        {\color{red}\neg(M\geq 0)\wedge  N\leq1000},\allowbreak
        {\color{red}M\geq 0\wedge \neg(N\leq1000)}\}$.
More details on how to implement MC/DC can be found in~\cite{mcdc}.

However, MC/DC is not immediately applicable for to software specifications,
as it does not support equivalence class partitioning or boundary
value analysis.  Nevertheless, we believe MC/DC generates a meaningful set
of testing conditions, which can be used as the starting point for
adequately covering a requirement.  Thus, subsequently to plain MC/DC, we
perform two additional steps to obtain a larger set of testing conditions.

The motivation behind this is that an ordered predicate typically represents
one equivalence class for the variables involved.
As an example, given a non-ordered predicate $M\geq 0$, by convention
there are two equivalence classes: $M>0$ and $M=0$. In the following, we
are going to introduce two operations that eliminate non-ordered predicates
in any Boolean expression, by converting it into a set of Boolean
expressions that do not contain non-ordered predicates.

\paragraph{Negation-free expressions} %Negation-free for Boolean predicates}
A Boolean expression is said to be negation free for its predicates if
there is no negation applied to any Boolean predicate within it.  As an
example, the two colored testing conditions in the set $\Phi$ above are not
negation free.

Suppose that $e$ is a negated predicate of the form $e= \neg(nbe_1\sim
nbe_2)$. We define a $neg\mbox{-}free$ operator, as in 
Equation \eqref{eq:neg-free}, to convert a negation expression to its
equivalent set of non-negation expressions.  By recursively applying it 
to any appearance of the negated predicate within an expression, the 
$neg\mbox{-}free$ operator can be naturally extended to apply to any
Boolean expression.

\begin{equation}
  \label{eq:neg-free}
  neg\mbox{-}free(e)=\left\{
  \begin{aligned}
  \{ nbe_1=nbe_2,  nbe_1 > nbe_2\}\mbox{ if }\sim\mbox{ is } <  \\
  \{ nbe_1 > nbe_2\}\mbox{ if }\sim\mbox{ is } \leq  \\
  \{ nbe_1<nbe_2,  nbe_1 > nbe_2\}\mbox{ if }\sim\mbox{ is } =  \\
  \{ nbe_1 < nbe_2\}\mbox{ if }\sim\mbox{ is } \geq  \\
  \{ nbe_1=nbe_2,  nbe_1 < nbe_2\}\mbox{ if }\sim\mbox{ is } >  \\
  \{ nbe_1 = nbe_2\}\mbox{ if }\sim\mbox{ is } \neq  \\
  \end{aligned}
  \right.
\end{equation}

For the aforementioned example in~\eqref{eq:mcdc-example}, if we apply the
negation free operator to its MC/DC results, the resulting testing
conditions are
$\Phi=\{M\geq 0\wedge N\leq1000,\allowbreak
        {\color{red} M<0}\wedge N\leq1000,\allowbreak
        M\geq 0\wedge {\color{red}N>1000}\}$.
The colored predicates are due to $neg\mbox{-}free$.

\paragraph{Non-ordered predicate expansion}
Given a non-ordered Boolean predicate $e=nbe_1\sim nbe_2$ with $\sim\in
\{\leq,\geq,\neq\}$, the $to\mbox{-}ordered$ operator is defined to expand
it into the equivalent ordered predicates.
\begin{equation}
  \label{eq:to-ordered}
  to\mbox{-}ordered(e)=\left\{
  \begin{aligned}
  \{ nbe_1 < nbe_2, nbe_1=nbe_2\}\mbox{ if }\sim\mbox{ is } \leq  \\
  \{ nbe_1 > nbe_2, nbe_1=nbe_2\}\mbox{ if }\sim\mbox{ is } \geq  \\
  \{ nbe_1 < nbe_2, nbe_1>nbe_2\}\mbox{ if }\sim\mbox{ is } \neq  \\
  \end{aligned}
  \right.
\end{equation}

Accordingly, for any testing condition generated after the MC/DC starting
point, by conducting the $neg\mbox{-}free$ and $to\mbox{-}ordered$
operations, we obtain an elaborated set of testing conditions that respect
equivalence class partitioning.  As a result, for the example
in~\eqref{eq:mcdc-example}, the final set of testing conditions will be
$\Phi=\{
M=0\wedge N=1000,\allowbreak\hspace{0.5ex}
M=0\wedge N<1000,\allowbreak\hspace{0.5ex}
M>0\wedge N=1000,\allowbreak\hspace{0.5ex}
M>0\wedge N<1000,\allowbreak \hspace{0.5ex}
M<0\wedge N=1000,\allowbreak\hspace{0.5ex}
M<0\wedge N<1000,\allowbreak\hspace{0.5ex}
M>0\wedge N>1000,\allowbreak\hspace{0.5ex}
M=0\wedge N>1000\}$.
Clearly, the final results in $\Phi$ combine these equivalence classes for
$M$ and $N$ based on their boundaries: in this case, $M=0$ and $N=1000$.

It is common to use a \emph{tolerance level} $\sigma$ for boundary value
analysis. The value of this tolerance level is based on the precision of the number
representation and requirements for the accuracy of calculation.  This
combines well with our coverage criterion.  In order to integrate this
tolerance level, we extend the $neg\mbox{-}free$ in~\eqref{eq:neg-free} and
$to\mbox{-}ordered$ in~\eqref{eq:to-ordered} such that every time a testing
condition with a predicate of the form $nbe_1<nbe_2$ (resp.~$nbe_1>nbe_2$)
is obtained, an extra testing condition with this predicate replaced by
$nbe_1=nbe_2-\sigma$ (resp.~$nbe_1=nbe_2+\sigma$) is added to the set.

\subsection{The ITE expression}
So far, the proposed coverage criterion extends the MC/DC rules with the boundary value analysis, but it has no consideration
of the possible control flow structure within a requirement, which is tackled in this part. %It is far too complex to analyze
%exhaustive control flows of these requirements, instead, our 
The focus is on the (nested) if-then-else style structures encoded in the requirement.
%case that a requirement encodes.
%
Each if-then-else (ITE) expression is represented as the tuple $ITE(e_1,
e_2, e_3)$, where $e_1$, $e_2$ and $e_3$ are all Boolean expressions such
that $e_2$ will be reached only if $e_1$ is true, otherwise $e_3$ is chosen. 
Note that if $e_2$ and $e_3$ are not Boolean, the coverage analysis would
stop at $e_1$.  An example of the ITE expression is
\begin{equation}
  \label{eq:ite-example}
  ITE(M+N\leq 10, Res=M+N, Res=10) \;.
\end{equation}

Given any Boolean expression $e$, we use $\Phi(e)$ to denote its set of
testing conditions after $neg\mbox{-}free$ and $to\mbox{-}ordered$
operations, following the MC/DC phase.  Besides, $\Phi(e,+)$ contains all
these testing conditions representing that the decision of $e$ is true and
$\Phi(e,-)$ is for $e$ to be false such that
$\Phi(e)=\Phi(e,+)\cup\Phi(e,-)$.

Regarding any if-then-else expression $e=ITE(e_1, e_2, e_3)$, the coverage
analysis of $e_2$ (resp.  $e_3$) will be only relevant if $e_1$ is true
(resp.  false).  Therefore, we define the resultant testing conditions
from an ITE expression $e$ as follows.
\begin{equation}
  \label{eq:ite-coverage}
  \begin{aligned}
    \Phi(e)=&\{e'|\exists x\in\Phi(e_1,+)\,\exists y\in\Phi(e_2)\mbox{ s.t. }e'=x\wedge y\}\\ 
  &\cup \{e'|\exists x\in\Phi(e_1,-)\,\exists y\in\Phi(e_3)\mbox{ s.t. }e'=x\wedge y\}
  \end{aligned}
\end{equation}

If we apply the definition in \eqref{eq:ite-coverage} to the ITE expression
in \eqref{eq:ite-example}, the final set of testing conditions generated
becomes
%\[\begin{array}{l@{}l}
%\Phi=\{ & M+N<10\wedge Res=M+N,~~{\color{red}M+N<10\wedge Res>M+N}, \\
%        & {\color{red}M+N<10\wedge Res<M+N},~~M+N=10\wedge Res=M+N, \\
%        & {\color{red}M+N=10\wedge Res>M+N},~~{\color{red}M+N=10\wedge Res<M+N}, \\
%        & M+N>10\wedge Res=10,~~{\color{red}M+N>10\wedge Res>10}, \\
%        & {\color{red}M+N>10\wedge Res<10} \}\;.
%\end{array}
%\]
\begin{equation}
  \label{eq:phi}
  \begin{array}{l@{}l}
    \Phi=\{ & M+N<10\wedge Res=M+N,~~{\color{red}M+N<10\wedge}\\
          & {\color{red} Res>M+N},~~{\color{red}M+N<10\wedge Res<M+N},\\
          & M+N=10\wedge Res=M+N,~~{\color{red}M+N=10\wedge}\\
          & {\color{red} Res>M+N},~~{\color{red}M+N=10\wedge Res<M+N}, \\
          & M+N>10\wedge Res=10,~~{\color{red}M+N>10\wedge Res}\\
          & {\color{red} >10},~~{\color{red}M+N>10\wedge Res<10} \}\;.
\end{array}
\end{equation}
It can be easily seen that if \eqref{eq:ite-example} is some requirement
that must be guaranteed in the program, these colored testing conditions
shall never be met.

\section{The Automated Testing Procedure}
\label{sec:auto-gen}

In this section, we give the procedure for automated test generation from a
function's requirements, which are specified as pre-conditions ($Pre$) and
post-conditions ($Post$). This is a generic framework to apply Bounded Model
Checking (BMC) for automatically creating test cases from a function's 
requirements, and it relies on several routine methods (e.g., non-deterministic
assignment) that are commonly available in modern Model Checkers. In our
particular case, we implement this automated testing inside CBMC \cite{cbmc}.

Suppose that these requirements are for a
function/procedure~$\mathcal{F}$, which has a set of input parameters $\mathcal{I}$
and a set of output parameters $\mathcal{O}$.  The function~$\mathcal{F}$
may also require some global input variables in $\mathcal{I}^g$ and after
the function call it can change the values of some global output variables
in $\mathcal{O}^g$.

In order to test the requirements of the target function $\mathcal{F}$, we
require a \emph{calling context} $\mathcal{F}^c$ for it.  The calling
context can be automatically generated, e.g., by a Python script.
A concrete example is available in the appendix.

The calling context function $\mathcal{F}^c$ is structured as in
Listing~\ref{list:calling_context}.  REQUIRE is supposed to be the method
that is supplied by a Model Checker.  It takes a Boolean expression as the
input argument and guarantees this expression is true/valid for the rest
part of the program.  It can also take string information as the second
input argument, which is necessary for the traceability of tests, however,
for simplicity this is not explicitly specified here.

\lstset{language=C,
basicstyle=\ttfamily,%\scriptsize,
frame=single,
keywordstyle=\color{blue}\ttfamily,
stringstyle=\color{red}\ttfamily,
commentstyle=\color{green}\ttfamily,
morecomment=[l][\color{magenta}]{\#}
}
\begin{lstlisting}[mathescape=true, 
caption={The calling context function}, 
captionpos=b,
label={list:calling_context}, 
%numbers=left,
commentstyle=\color{red}\ttfamily]
void $\mathcal{F}^c$(void)
{
  /** input type check */
  $\forall$ var $\in \mathcal{I}\cup\mathcal{I}^g$
    var := IN_TYPE(var_type_expr);

  /** preconditions */
  $\forall$ expr $\in Pre$
    REQUIRE(expr)

  /** over-approximation call of $\mathcal{F}$ */

  /** output type check */
  $\forall$ var $\in \mathcal{O}\cup\mathcal{O}^g$
    var := IN_TYPE(var_type_expr);

  /** postconditions */
  $\forall$ expr $\in Post$
    REQUIRE(expr)
}
\end{lstlisting}

Initially, all inputs of the function $\mathcal{F}$ are
non-deterministically assigned a value within the valid range of its type,
by the IN\_TYPE method, which contains two steps:
\begin{enumerate}
\item to generate a non-deterministic value for $var$;
\item to guarantee the value at step (1) is valid subject to $var\_type\_expr$ 
       and this is achieved by calling the REQUIRE method.
\end{enumerate}

For any variable $var$, the $var\_type\_expr$ is a Boolean expression
depends on the type of $var$.  For example,
\begin{itemize}
  \item if $var$ is a float variable, then $var\_type\_expr$ is in form of $min\_float\leq var\leq max\_float$;
  \item if $var$ is of an enum type, then $var\_type\_expr$ is in form of $var=enum_1\vee var=enum_2\vee\ldots$,
  and it enumerates all possible values of $var$.
  %where $enum_1$, $enum_2$ et al are possible values of the corresponding enum type.
\end{itemize}
As a matter of fact, these input checks cope with the implicit requirements of software.
That is, every input/output variable in the program inherently is constrained by its type.

Subsequently, for every requirement in the pre-conditions ($Pre$), a call to
the REQUIRE method guarantees that it must be true for the following part of
the program context.  Because we aim to independently examine the
requirements specified as pre-/post-conditions, the function body of
$\mathcal{F}$ is simply ignored, whereas a proper over-approximation is also
allowed depending on the application scenario.  In consequence, outputs from
$\mathcal{F}$ are over-approximately assigned values according to their
types.  To keep the consistency, requirements from the post-conditions must
be required to be true too, which indicates that our method does not target
these conditions out of the range of the requirements.

In order to conduct the automated testing procedure, coverage criterion in
Section \ref{sec:criterion} is supposed to have been already implemented in a Model 
Checker that is CBMC in this work. There are a number of works that implement or integrate
different kinds of coverage analyses with Bounded Model Checking. Examples are like
\cite{Ball2005}, \cite{angeletti2010using} and \cite{suman2010masking}.
We have adopted a similar approach that embeds trap properties 
(i.e., negation of testing conditions) into the program model of these
requirements.

The Model Checker simply goes through
the calling context function $\mathcal{F}^c$ as listed % \ref{list:calling_context} 
and generates the set of testing conditions $\Phi$ (e.g., Equation (\ref{eq:phi})) for
every $expr$ and $var\_type\_expr$, according to the coverage criterion defined.

For each testing condition (let us  say) $\phi\in\Phi$, a trap property (that violates
this testing condition) is encoded as $\neg\phi$ and instrumented into the $\mathcal{F}^c$
context, after which the Model Checker checks if these trap properties are
satisfied.
\begin{itemize}
\item In case a (trap) property is not satisfied, then an execution trace of
the program model (of $\mathcal{F}^c$) leading to the reachability of the
violation of the property (i.e., the original testing condition) is returned
and a test vector is outputted.
\item Otherwise, it means that the corresponding testing condition will never be met.
\end{itemize}

In conclusion, a test vector in the results represents an admissible
behaviour with respect to the function requirements, and it is expected that
given the inputs as in a test vector, the execution of the function
implementation shall trigger the corresponding testing conditions for
examining the requirements, but it is not necessarily true that the exact
values of outputs must be always the same, depending on the implementation
choices.

%It is worth noting that each test vector in the final results test against certain testing conditions within
%some pre or post conditions
%Applying the coverage criterion defined in Section xx to generate trap conditions...

%\input{exec-gen}

\section{The Case Study}
\label{sec:case-study}

The functional requirements-based test-generation method introduced in this
paper has been implemented in the Model Checking tool CBMC,
and is integrated into the toolchain developed in the AUTOSAC (Automated
Testing of SPARK Ada Contracts) project.

The AUTOSAC toolchain is encapsulated within RapiTestFramework,
the unit- and system-level testing solution provided by Rapita Systems. 
It serves as the graphical environment presented to the user as well as the
solution for compiling, executing, and reviewing the generated tests.
This toolchain's performance and ability to replace manual testing effort is
confirmed through the process of generating and executing tests for selected
code made available by Rolls-Royce.

\subsection{Test compilation, execution and review}
\label{sec:exec-gen}
As depicted in Figure \ref{fig:scenario}, DO-178C certification requests
evidence from both requirements coverage analysis and structural coverage
analysis. This means, beyond automatically generating tests from requirements,
a successful testing toolchain also needs to provide the following:

\begin{itemize}
   \item executes the procedure or function under test with the test vector inputs;
   \item collects coverage data for execution of the subprogram, including statement, branch, and/or MC/DC coverage information;
   \item verifies that the Post-condition(s) for the subprogram have not been violated when the subprogram exits.
\end{itemize}

In addition, the data resulting from one or more test runs must be analyzed, collated, and presented to the tester in a format suitable for review. In this context, review may include either checking the successful execution of the test(s) or reviewing the percentage of code coverage acheived during execution.

RapiTestFramework provides an existing solution for unit test execution and code coverage reporting, and has been extended in the following ways to acheive to achieve these goals:

\begin{itemize}
   \item a dialog is presented to the tester displaying available subprograms for testing;
   \item a documented XML schema for test vector definitions has been exposed in order to allow communication with external tools such as CBMC;
   \item the generated test harness includes exception-handling for the purpose of reporting violated post-conditions.
\end{itemize}

The coverage reporting features of RapiTestFramework have been found to be suitable for review of test execution, especially code coverage percentages and source code highlighting and annotation.

\subsection{Overview of toolchain workflow}

The AUTOSAC case study required the integration of CBMC for test vector
generation and RapiTestFramework for subprogram selection, test compilation
and execution, and review of results.

%In addition to demonstrating the efficacy of the approach, the case study required consideration of some other factors. These include:
%\begin{itemize}
%   \item the practicalities of generating executable tests;
%   \item an convenient interface for selecting subprograms to test and initiating the process;
%   \item providing useful feedback to the user during generation and execution;
%   \item displaying test results and code coverage post-execution for review;
%\end{itemize}

This process is managed using a graphical environment to call the underlying tools.
A brief explanation of each stage in the testing workflow follows,
beginning with software requirements and concluding with test execution results.

The starting point for test generation is the low-level requirements of the
functions and procedures under test, formulated as SPARK pre- and
post-conditions.  Because the generated tests are based on these
requirements, they have the same level of independence from software
implementation as with a manual test-writing process.

Firstly, these requirements are analyzed (by CBMC). Pre-/post-conditions and ranges
of input/output parameter types define the scope of possible test vectors,
and they are analyzed using strategies discussed in this paper for the test
generation.
%
%Post-condition(s) are then analyzed to select a set of meaningful vectors 
%using strategies such as boundary-value analysis and equivalence class partitioning as discussed in this paper. 
%
To the extent to which these algorithms are successful, the collection of
test vectors will be of similar quantity and quality to those that might
have been written by a test engineer.  The resulting test vectors are stored
in a documented XML format.

\begin{figure}[!htb]
    \centering
    \includegraphics[width=0.85\textwidth]{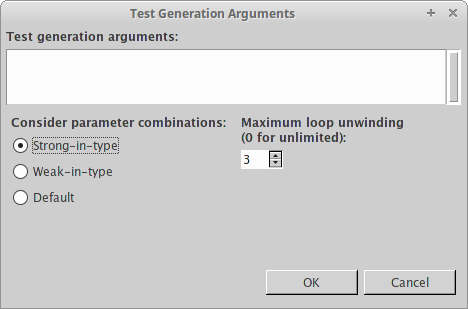}    
    \caption{Toolchain configuration options}
    \label{fig:screenshot1}
\end{figure}

Some configurable options were explored during the course of the case study
which affect the resultant generated vectors, including limiting the
unwinding of loops in the analysis of SPARK contracts, and the exploration
of different combination levels of in-typed parameters for the coverage
analysis, as in Figure \ref{fig:screenshot1}.

%the exploration of Cartesian/strong or pair-wise/weak product combinations of in-typed parameters to a subprogram. These options have been exposed to the user in a dialog. %

\begin{figure}[!htb]
    \centering
    \includegraphics[width=0.85\textwidth]{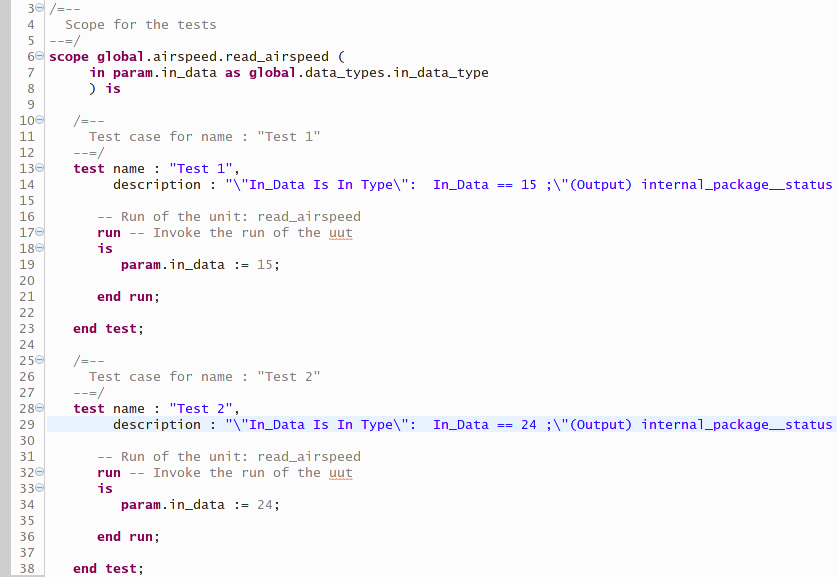}    
    \caption{The test script}
    \label{fig:screenshot2}
\end{figure}

The next stage of the process converts these test vectors into executable
test scripts (as in Figure~\ref{fig:screenshot2}).  The scripts are
presented to the user in their test project and are the principle artifact
of the generation process.  These tests may be altered by hand in the
future, or simply used for test review; to facilitate this, the interface
provides a text editor with syntax highlighting and error checking.  Note
that the testing condition is specified as part of the \emph{description},
which provides the traceability back to requirements.

\begin{figure}[!htb]
    \centering
    \includegraphics[width=0.85\textwidth]{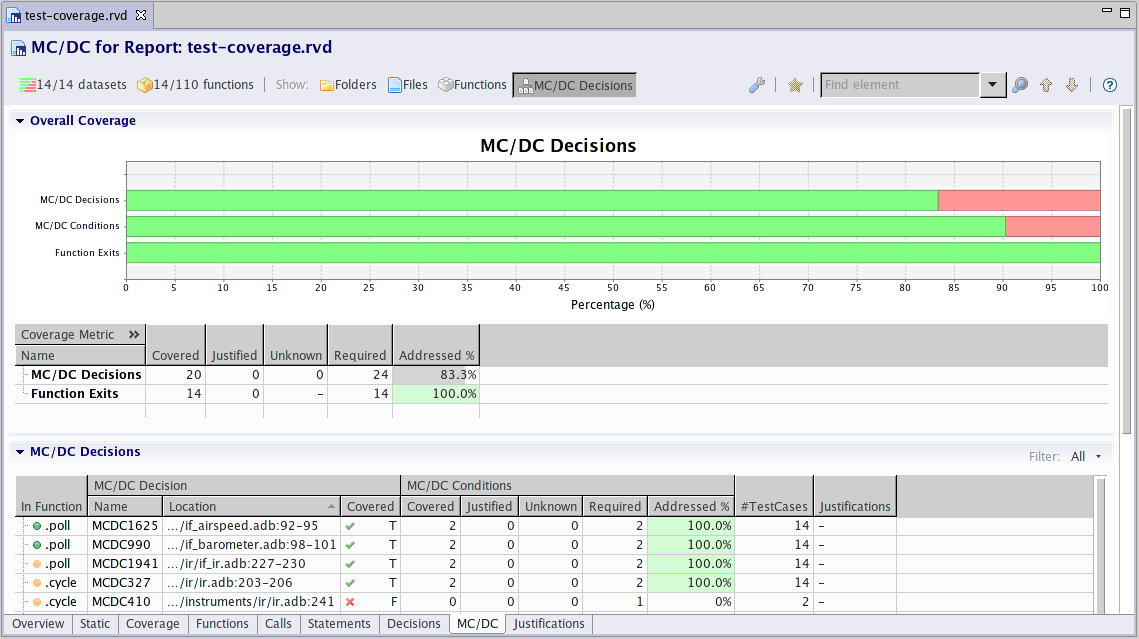}    
    \caption{The code coverage analysis}
    \label{fig:screenshot3}
\end{figure}

Finally, the GUI interface enables the user to compile and run a test
executable including some or all of the generated tests.  Upon completion of
the test execution a report is created, which contains both test and code
coverage results for analysis and review.  Users can view coverage metrics
at the function, statement, or MC/DC level, providing immediate feedback on
both the effectiveness of test generation as well as the coverage of the
source code achieved from requirements, as in Figure~\ref{fig:screenshot3}. 
The coverage reporting features of RapiTestFramework, especially code
coverage percentages and source code highlighting and annotation, have been
found to be suitable for review of test execution.

In addition to the material presented here, a video demonstration of the
toolchain (using open-source code under test) is available
online\footnote{https://youtu.be/72NWFZQOvlM}.

%The case study undertaken during the AUTOSAC research project demonstrated the successful implementation of automated testing in a real-world environment, integrating and extending existing tools for achieving DO-178-level certification. The tools required for this have been bundled (together with documentation) in an installer which was distributed to Rolls-Royce as a functional prototype. We anticipate that the toolchain will undergo further development, including adding support for mixed-language test projects.

\subsection{Case study results}

The software architecture behind the case studies is composed of two layers. 
The first one is an application layer (AL) and the second one, is an
operating layer (OL).  In order to use the toolchain in a real environment,
two case studies have been picked, one per each layer in order to observe
how the tool works with different abstract levels.  Concretely, a health
monitoring component for the AL case study and a temperature sensor
component for the OL case study, both written in Ada, have been chosen.

After both components have been isolated from non-related code, to make the
toolchain work, the legacy code has to be modified to represent the
requirements with the component/functions by the use of SPARK Ada contracts,
specifically, the use of SPARK pre- and post- conditions.

Ghost functions from SPARK2014 have not been used, as they are not currently
supported by the toolchain.  This has not been a problem for proceeding with
the examples, though this would be a great advantage for the future.

By applying the AUTOSAC toolchain, tests are automatically generated from
the SPARK Ada contracts.  We check these resultant tests, by measuring the
corresponding statement code coverage level on each subprogram:
$AFA (100\%)$, $AST (100\%)$, $CTE1 (100\%)$, $CTE2 (100\%)$, $CTEC (100\%)$, $CVO (100\%)$, $FES$ $(100\%)$, $ISCD (100\%)$, $ISCF (100\%)$, $ISCC (100\%)$,  $RTC (68.75\%)$.

Overall, the automatically generated tests show good coverage and represent
a good spectrum of the case scenarios to be tested.  The gap on code
coverage level in some cases (e.g., $RTC$ from the OL) required further
refinements on the SPARK Ada contracts under test and investigations on the
intermediate language representation for SPARK Ada contracts in the
toolchain.

Importantly, the verification role will retain its skillset, yet remove
monotonous test scripting, as a result of writing SPARK Ada contracts
instead of specific test harness.  Besides writing such contracts, we
believe minimal human observation for the review of the automatically
generated tests is required.

%In the end, the AUTOSAC toolchain shows the great promise and high potential within  the internal processes of Rolls-Royce and has become one of the candidates for Rolls-Royce to reduce costs for verification, which at the moment is of paramount expense.
The case study undertaken during the AUTOSAC research project demonstrated the successful implementation of automated testing in a real-world environment, integrating and extending existing tools for achieving DO-178-level certification. The toolchain has become one of the candidates for Rolls-Royce to reduce costs for verification, which is of great interest. We anticipate that it will undergo further development, including adding support for mixed-language test projects.

\subsection{Experience and lessons learned}

The main obstacle we encountered during the toolchain development phase is
the intermediate representation of SPARK Ada.  Regarding the automated
testing procedure proposed in this paper, its implementation is based on
CBMC, which does not have a direct front end for Ada.  Thus, these
requirements specified as in SPARK Ada contracts are at first converted into
C language form, from which tests are then automatically generated.  This
choice is due to the limited timescale of the AUTOSAC project and is also
for the purpose of evaluating the feasibility of such a requirements-based
automated testing methodology.  However, certain features in Ada cannot be
trivially represented in C and the conversion procedure from Ada to C needs
to be very carefully maintained.  Currently, there is a continuing
collaboration between partners to develop a formal Ada/SPARK front end for
CBMC, thanks to the promising results shown from AUTOSAC toolchain
for the automated test generation from software requirements.

Notwithstanding this limitation, the case study presented here illustrates
the success of the tools and approach developed during the AUTOSAC project. 
In particular, the successful integration of CBMC and RapiTestFramework
demonstrated the powerful combination of test generation based on
mathematical proof combined with a robust platform for test execution and
review.  It is expected the reduction in time and cost of low-level
requirement validation described in this paper will begin to be achieved as
this technology is adopted in large-scale commercial projects.

On the other hand, defining requirements in the SPARK
language has shown four advantages.  Firstly, the different abstract levels
allow us to generate low level tests, but also high level tests.  Secondly,
SPARK contracts provide an additional implicit verification between
requirements and code design.  Thirdly, it allows legacy code to become
closer to new formal methods, as the evolution of SPARK language, so in
future, the change would not be very radical.  Finally, the use of the
AUTOSAC toolchain encourages software teams to adopt agile development
practices, as quick tests of the code can be done while maintaining
independence from implementation, something that is currently not possible
within the constraints of the DO-178 guidance.

Regarding this last point, such an agile approach allows the same person to
perform two roles on different modules. The first one is the verification role,
which would initially make the specification of some modules using SPARK Ada contracts. 
The other one is the design role, which would design/implement a different
software module, independently from the SPARK Ada contracts written by
another person.

\section{Conclusions}
\label{sec:conclusions}

In this paper, we successfully demonstrated feasibility of automatic test case
generation from functional requirements, targeting software testing in avionics.
We define a requirements-based coverage criterion and formulate an automated
procedure for functional requirements-based testing. We implemented our method
and integrated it within a testing environment. The developed toolchain is
applied to industrial case studies, and its applicability and usefulness are
confirmed. 

Regarding future exploitation, we are interested in investigating application
of the technique in this paper to test case chains generation for reactive 
systems \cite{smk2014}. In particular, testing of model-based development processes
(e.g., testing Simulink systems~\cite{fmco2009,hrk2011-dac,ale2017sies}) is of great
relevance to the context of avionics software certification.

\section*{Acknowledgement}
The work presented in this paper was sponsored by Birmingham City Council
and Finance Birmingham and supported by Farnborough Aerospace Consortium
(FAC) under NATEP project FAC043 -- Automated Testing of SPARK Ada Contracts
(AUTOSAC).

\let\oldbibliography\thebibliography
\renewcommand{\thebibliography}[1]{\oldbibliography{#1}
\setlength{\itemsep}{2pt}}

\bibliographystyle{IEEEtran}
\bibliography{all}

\begin{thebibliography}{10}
\providecommand{\url}[1]{#1}
\csname url@samestyle\endcsname
\providecommand{\newblock}{\relax}
\providecommand{\bibinfo}[2]{#2}
\providecommand{\BIBentrySTDinterwordspacing}{\spaceskip=0pt\relax}
\providecommand{\BIBentryALTinterwordstretchfactor}{4}
\providecommand{\BIBentryALTinterwordspacing}{\spaceskip=\fontdimen2\font plus
\BIBentryALTinterwordstretchfactor\fontdimen3\font minus
  \fontdimen4\font\relax}
\providecommand{\BIBforeignlanguage}[2]{{%
\expandafter\ifx\csname l@#1\endcsname\relax
\typeout{** WARNING: IEEEtran.bst: No hyphenation pattern has been}%
\typeout{** loaded for the language `#1'. Using the pattern for}%
\typeout{** the default language instead.}%
\else
\language=\csname l@#1\endcsname
\fi
#2}}
\providecommand{\BIBdecl}{\relax}
\BIBdecl

\bibitem{do178c}
RTCA, ``{DO-178C}, {S}oftware considerations in airborne systems and equipment
  certification,'' 2011.

\bibitem{bs7925}
S.~C. Reid, ``{BS 7925-2}: The software component testing standard,'' in
  \emph{Quality Software. Proceedings. First Asia-Pacific Conference on}.\hskip
  1em plus 0.5em minus 0.4em\relax IEEE, 2000, pp. 139--148.

\bibitem{gargantini1999using}
A.~Gargantini and C.~Heitmeyer, ``Using model checking to generate tests from
  requirements specifications,'' in \emph{Software
  Engineering—ESEC/FSE}.\hskip 1em plus 0.5em minus 0.4em\relax Springer,
  1999, pp. 146--162.

\bibitem{chang1999structural}
J.~Chang and D.~J. Richardson, ``Structural specification-based testing:
  {a}utomated support and experimental evaluation,'' in \emph{Software
  Engineering—ESEC/FSE}.\hskip 1em plus 0.5em minus 0.4em\relax Springer,
  1999, pp. 285--302.

\bibitem{hierons2009using}
R.~M. Hierons, K.~Bogdanov, J.~P. Bowen, R.~Cleaveland, J.~Derrick, J.~Dick,
  M.~Gheorghe, M.~Harman, K.~Kapoor, P.~Krause \emph{et~al.}, ``Using formal
  specifications to support testing,'' \emph{ACM Computing Surveys (CSUR)},
  vol.~41, no.~2, p.~9, 2009.

\bibitem{tan2004specification}
L.~Tan, O.~Sokolsky, and I.~Lee, ``Specification-based testing with linear
  temporal logic,'' in \emph{Information Reuse and Integration. Proceedings of
  the IEEE International Conference on}, 2004, pp. 493--498.

\bibitem{whalen2006coverage}
M.~W. Whalen, A.~Rajan, M.~P. Heimdahl, and S.~P. Miller, ``Coverage metrics
  for requirements-based testing,'' in \emph{Proceedings of the international
  symposium on Software testing and analysis}.\hskip 1em plus 0.5em minus
  0.4em\relax ACM, 2006, pp. 25--36.

\bibitem{cbmc}
E.~Clarke, D.~Kroening, and F.~Lerda, ``A tool for checking {ANSI-C}
  programs,'' in \emph{International Conference on Tools and Algorithms for the
  Construction and Analysis of Systems}.\hskip 1em plus 0.5em minus 0.4em\relax
  Springer, 2004, pp. 168--176.

\bibitem{do333}
RTCA, ``{DO-333}, {Formal} methods supplement to {DO-178C} and {DO-278A},''
  2011.

\bibitem{delahaye2013common}
M.~Delahaye, N.~Kosmatov, and J.~Signoles, ``Common specification language for
  static and dynamic analysis of {C} programs,'' in \emph{Proceedings of the
  28th Annual ACM Symposium on Applied Computing}.\hskip 1em plus 0.5em minus
  0.4em\relax ACM, 2013, pp. 1230--1235.

\bibitem{barnett2004spec}
M.~Barnett, K.~R.~M. Leino, and W.~Schulte, ``The {S}pec\# programming system:
  {an} overview,'' in \emph{International Workshop on Construction and Analysis
  of Safe, Secure, and Interoperable Smart Devices}.\hskip 1em plus 0.5em minus
  0.4em\relax Springer, 2004, pp. 49--69.

\bibitem{spark}
B.~John, ``{SPARK}: The proven approach to high integrity software,'' 2012.

\bibitem{chilenski2001investigation}
J.~J. Chilenski, ``An investigation of three forms of the modified condition
  decision coverage ({MCDC}) criterion,'' DTIC Document, Tech. Rep., 2001.

\bibitem{mcdc}
K.~J. Hayhurst, D.~S. Veerhusen, J.~J. Chilenski, and L.~K. Rierson, ``A
  practical tutorial on modified condition/decision coverage,'' 2001.

\bibitem{Ball2005}
T.~Ball, \emph{A Theory of Predicate-Complete Test Coverage and
  Generation}.\hskip 1em plus 0.5em minus 0.4em\relax Berlin, Heidelberg:
  Springer Berlin Heidelberg, 2005, pp. 1--22.

\bibitem{angeletti2010using}
D.~Angeletti, E.~Giunchiglia, M.~Narizzano, A.~Puddu, and S.~Sabina, ``Using
  bounded model checking for coverage analysis of safety-critical software in
  an industrial setting,'' \emph{J. Autom. Reasoning}, vol.~45, no.~4, pp.
  397--414, 2010.

\bibitem{suman2010masking}
P.~V. Suman, T.~Muske, P.~Bokil, U.~Shrotri, and R.~Venkatesh, ``Masking
  boundary value coverage: effectiveness and efficiency,'' in
  \emph{Testing--Practice and Research Techniques}.\hskip 1em plus 0.5em minus
  0.4em\relax Springer, 2010, pp. 8--22.

\bibitem{smk2014}
P.~Schrammel, T.~Melham, and D.~Kroening, ``Generating test case chains for
  reactive systems,'' \emph{Software Tools for Technology Transfer (STTT)},
  vol.~18, pp. 319--334, June 2016.

\bibitem{fmco2009}
A.~Brillout, N.~He, M.~Mazzucchi, D.~Kroening, M.~Purandare, P.~R{\"u}mmer, and
  G.~Weissenbacher, ``Mutation-based test case generation for {Simulink}
  models,'' in \emph{Formal Methods for Components and Objects (FMCO) 2009},
  ser. LNCS, vol. 6286.\hskip 1em plus 0.5em minus 0.4em\relax Springer, 2010,
  pp. 208--227.

\bibitem{hrk2011-dac}
N.~He, P.~R{\"u}mmer, and D.~Kroening, ``Test-case generation for embedded
  {Simulink} via formal concept analysis,'' in \emph{Design Automation
  Conference (DAC)}.\hskip 1em plus 0.5em minus 0.4em\relax ACM, 2011, pp.
  224--229.

\bibitem{ale2017sies}
A.~Balsini, M.~Di~Natale, M.~Celia, and V.~Tsachouridis, ``Generation of
  {Simulink} monitors for control applications from formal requirements,'' in
  \emph{Industrial Embedded Systems (SIES), 12th IEEE Symposium on}, 2017.

\end{thebibliography}

\newpage
{\center\section*{Appendix}}
This is an illustrative example for the calling context
function presented in Section \ref{sec:auto-gen}.
Given a target function $\mathcal{F}$ named as 
\emph{constrained\_add}, there are two input
arguments $M$ and $N$, and one output $Res$.
The function's requirements are specified as 

\[\begin{array}{l@{}l}
\{&M\geq0\wedge M\leq 10,~~N\geq0\wedge N\leq 10,~~Res\geq 0\wedge Res\leq\\
  &10,~~ITE(M+N\leq 10, Res=M+N, Res=10)\}
\end{array}
\]

Using the AUTOSAC toolchain, the calling context function
$\mathcal{F}^c$ is generated as in the code List \ref{list:constrained_add}.
The two steps of IN\_TYPE method in Section \ref{sec:auto-gen} correspond to
the assignment of \emph{nondet\_\_my\_range} and the call to \emph{\_\_AUTOSAC\_in\_type}.

As mentioned in Section \ref{sec:auto-gen}, annotations can be also added
as strings. For example, "postcondition at basic\_tests.ads:52:20" tells
that this is a post-condition that can be traced back to line 52 at the 
source file basic\_tests.ads.

\lstset{language=C,
basicstyle=\ttfamily\scriptsize,
frame=single,
keywordstyle=\color{blue}\ttfamily,
stringstyle=\color{red}\ttfamily,
commentstyle=\color{green}\ttfamily,
morecomment=[l][\color{magenta}]{\#}
}
\begin{lstlisting}[mathescape=true, 
caption={The calling context function for \emph{constrained\_add}}, 
captionpos=b,
label={list:constrained_add}, 
commentstyle=\color{red}\ttfamily]
void ___calling_context__constrained_add () {
  /* in-parameters */
  my_range M;
  M = nondet__my_range();
  __AUTOSAC_in_type((M >= 0) && (M <= 10), "M is in type");
  my_range N;
  N = nondet__my_range();
  __AUTOSAC_in_type((N >= 0) && (N <= 10), "N is in type");
  my_range res;

  /* subprogram in-globals */

  /* subprogram preconditions */


  /* overapproximate call: out-parameters and out-globals */
  res = nondet__my_range();
  __AUTOSAC_in_type((Res >= 0) && (Res <= 10),
                    "Res is in type");

  /* explicit postcondition */
  __AUTOSAC_postcondition(((m + n) > 10) ? (res == 10) : 
                  (res == (m + n)),
                  "postcondition at basic_tests.ads:52:20");
}
\end{lstlisting}

A calling context function in such a form can be now directly recognized and
run through by CBMC. A set of testing conditions for each requirement of 
the \emph{constrained\_add} method will be generated according to the criterion
in Section \ref{sec:criterion}. As an example, for the post-condition
$ITE(M+N\leq 10, Res=M+N, Res=10)$, the set of testing conditions is as in
\eqref{eq:phi}. Afterwards, these conditions are negated and instrumented by CBMC
to automatically generate test cases. The whole procedure of test generation
is independent from the implementation of \emph{constrained\_add} function.

\end{document}